\documentclass[journal=aamick,manuscript=article]{achemso}
\setkeys{acs}{articletitle = true}
\usepackage{soul}
\usepackage{tabularx}
\usepackage[version=3]{mhchem} 
\usepackage[T1]{fontenc}       
\usepackage{epstopdf}
\DeclareGraphicsExtensions{.pdf,.eps,.png,.jpg,}
 
\usepackage{epstopdf}
\usepackage{multirow}
\usepackage{soul}
\usepackage[normalem]{ulem}
\usepackage{float}
\usepackage[utf8]{inputenc}
\usepackage[T1]{fontenc}
\usepackage{mathptmx}
\usepackage{etoolbox}
\usepackage{gensymb}
\usepackage{amsmath}
\usepackage{xcolor}
\usepackage{natbib}
\usepackage{alphabeta}
\usepackage{placeins}
\DeclareUnicodeCharacter{2009}{\,}
\usepackage[utf8]{inputenc}
\usepackage{hyperref,url}

\makeatletter
\def\@email#1#2{%
 \endgroup
 \patchcmd{\titleblock@produce}   {\frontmatter@RRAPformat}
  {\frontmatter@RRAPformat{\produce@RRAP{*#1\href{}{#2}}}\frontmatter@RRAPformat}
  {}{}
}%
\makeatother
  
\title{ Electric field management in $\beta$ – Ga$_2$O$_3$ vertical Schottky diodes using high-k bismuth zinc niobium oxide}
\author{Pooja Sharma*}
\author{Yeshwanth Parasubotu}
 \author{Saurabh Lodha}
\email{s.pooja.iitb@gmail.com;   slodha@ee.iitb.ac.in}
\affiliation{ Electrical Engineering Department, IIT Bombay, Mumbai, India - 400076
}%
\begin{document}\date{\today}

\begin{abstract}

In this work, we have integrated bismuth zinc niobium oxide (BZN), a high-k dielectric material, in metal-insulator-semiconductor (MIS) and field-plated metal-semiconductor (FP-MS) Schottky barrier diodes on $\beta$-Ga$_2$O$_3$. This increases the breakdown voltage (\textit{$V_{BR}$}) from 300 V to 600 V by redistributing the electric fields, leveraging the high permittivity of BZN (k $\sim$210). Enhancement in Schottky barrier height, by approximately 0.14 eV for MIS and 0.28 eV for FP-MS devices, also contributes to the improved $V_{BR}$. BZN inclusion has minimal impact on specific on-resistance ($R_{on,sp}$). Additionally, the devices display excellent current-voltage characteristics with ideality factors close to unity and an on/off current ratio greater than 10$^{10}$. This work presents the most significant $V_{BR}$ enhancement reported-to-date for MIS devices on $\beta$-Ga$_2$O$_3$ without compromising turn-on voltage and $R_{on,sp}$. A comparison of FP-MS and MIS devices shows that FP-MS outperforms MIS in terms of lower $R_{on,sp}$, higher Schottky barrier height, and improved $V_{BR}$.

\end{abstract}

\maketitle
\section*{\textbf{Introduction}}
Increasing demand for efficient power devices has led to significant interest in $\beta$-Ga$_2$O$_3$ devices due to its exceptional material properties. $\beta$-Ga$_2$O$_3$ is a wide bandgap semiconductor with a bandgap of 4.8 eV and a high theoretical breakdown field of  8 MV/cm. These characteristics make $\beta$-Ga$_2$O$_3$ a promising candidate for high-power and high-frequency applications. However, optimizing the breakdown characteristics of $\beta$-Ga$_2$O$_3$ diodes remains challenging due to the lack of a suitable shallow acceptor-type dopant, which limits the achievable barrier height to less than half of the bandgap. Additionally, high electric field concentration at the junction and device edges in Schottky diodes further complicates achieving high breakdown voltages.

Several techniques have been explored to improve the breakdown voltage ($V_{BR}$) and overall performance of $\beta$-Ga$_2$O$_3$ diodes. These include surface treatments \cite{psharmaAPL2024, JYang2018}, the use of interlayer dielectrics (ILs) such as SiO$_2 $\cite{bhattacharyya2020schottky}, Al$_2$O$_3$ \cite{Prajapati2024}, and boron nitride \cite{xu2023_BN} to enhance the barrier height and hence the interface electric field. However, the breakdown performance of these low-k dielectrics is limited by breakdown at the metal–dielectric interface, particularly near the device edge where the electric field peaks under reverse bias.
Heterostructures with p-type materials like NiO \cite{Lu2022_NiO, Hao2022_NiO} and Cu$_2$O \cite{Watahiki2017_Cu2O} have also been reported with enhanced $V_{BR}$. But NiO suffers with very low hole mobility\cite{sun2018_NiO_mobility, alidoust2015_NiO_mobility} ($<$ 10 cm$^2$/Vs) and Cu$_2$O is unstable, easily oxidising to form n-type CuO. 

In addition to these techniques, integrating high permittivity (k) dielectrics has proven to be particularly effective for enhancing the breakdown performance of $\beta$-Ga$_2$O$_3$ diodes. High-k dielectrics offer better electric field modulation and reduced leakage currents, making them ideal for high-power applications requiring large $V_{BR}$. While low-k dielectrics like SiO$_2$ or Al$_2$O$_3$ can provide high breakdown fields, they have higher internal electric fields, leading to breakdown at lower voltages. In contrast, high-k dielectrics distribute the electric field more efficiently along with lower internal electric fields, reducing the risk of dielectric breakdown.

High-k dielectrics, such as Nb$_2$O$_5$ (k $\sim$50) \cite{tiwari2021nb2o5} and BaTiO$_3$ (k $\sim$260)  \cite{roy2021high, xia2019metal}, have been explored for field management in various Schottky diode architectures, including metal-insulator-semiconductor (MIS) \cite{tiwari2021nb2o5, xia2019metal}, field-plated metal-semiconductor (FP-MS)\cite{roy2021high} and trench diodes\cite{roy2023BTO}, resulting in improved $V_{BR}$. While BTO integration has shown significant improvement in $V_{BR}$, it requires annealing for long duration ($\sim$30 minutes) and/or deposition at high temperatures, around 700°C  \cite{roy2023_BTO2, rahman2021_BTO, razzak2020_BTO, rahman2021_BTO2}. Although $\beta$-Ga$_2$O$_3$ can withstand high temperature processing,
it poses a significant challenge for co-integration with other materials, such as metal contacts and passivation layers, which may degrade or lose functionality at elevated temperatures. This limits the broader applicability of BTO in multi-material device architectures and underscores the need to identify high-k dielectric materials that provide both performance benefits and compatibility with standard semiconductor fabrication processes.
Premature breakdown in $\beta$-Ga$_2$O$_3$ diodes can result from edge field crowding, interface states, and dielectric breakdown in MIS structures with low-k materials. To mitigate these effects, high-k dielectrics can be used both, as field plates in FP-MS configuration as well as the IL in MIS SBDs.

\FloatBarrier
\begin{figure}[h]
    \centering
    \includegraphics[width=0.5\linewidth]{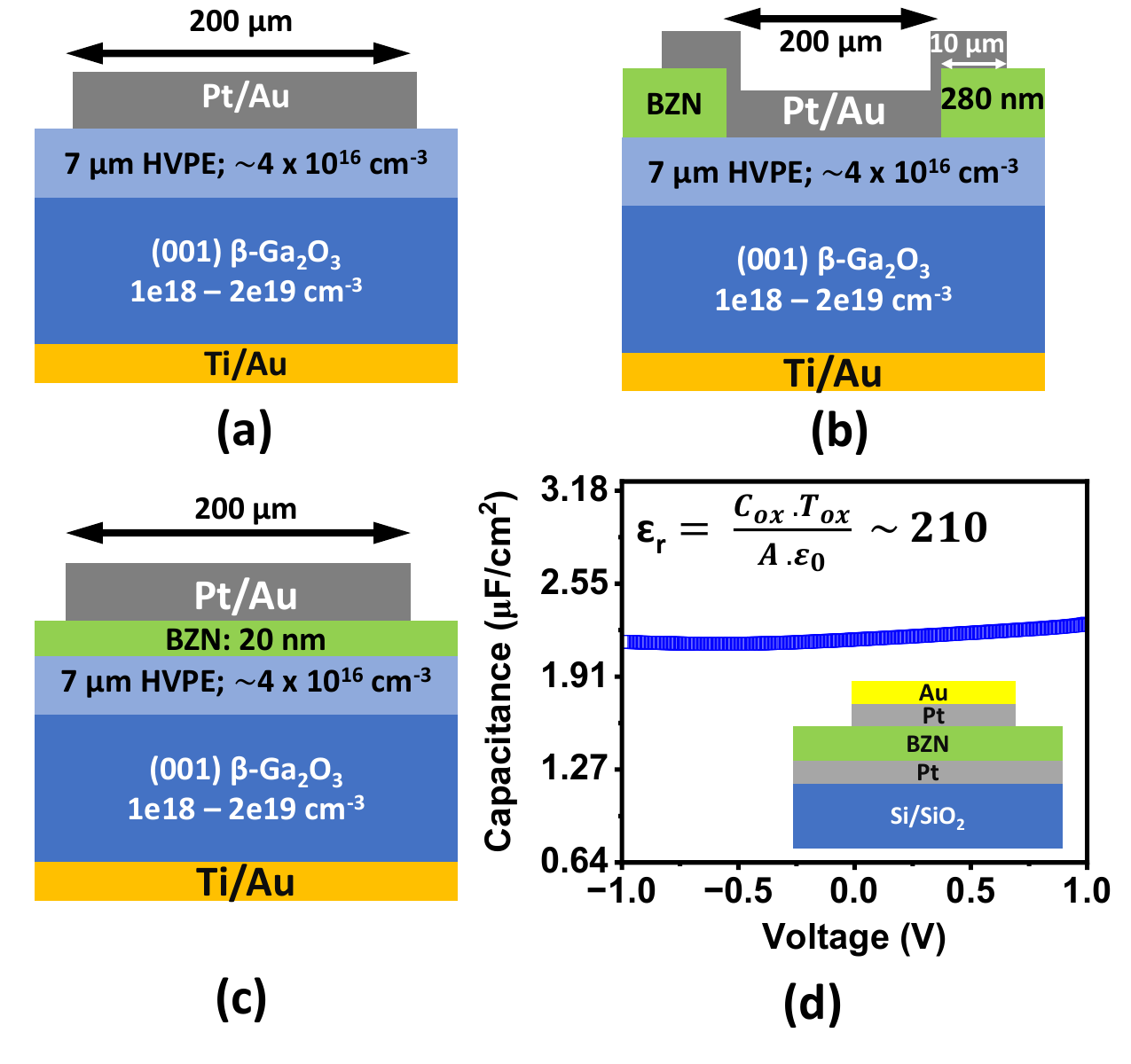}
    \caption{Device cross-section of 200 $\mu$m diameter (a) metal-semiconductor (MS), (b) field-plated metal-semiconductor (FP-MS), (c) metal-insulator-semiconductor (MIS) Schottky barrier diodes fabricated on 4 x 10$^{16}$ cm$^{-3}$ doped, 7 $\mu$m thick, $\beta$-Ga$_2$O$_3$ drift layers deposited using HVPE on conducting  $\beta$-Ga$_2$O$_3$ substrates with high-k dielectric BZN integration, and (d) capacitance-voltage characteristics of metal-BZN-metal capacitor (cross-section shown in inset) showing extraction of dielectric constant of BZN.  }
    \label{fig:fig0}
\end{figure}
\FloatBarrier
In this work, we have used low temperature (350 $\degree$C) deposited bismuth zinc niobium oxide (BZN), a high-k dielectric with dielectric constant value of nearly 210, to fabricate FP-MS and MIS Schottky barrier diodes (SBDs) on $\beta$-Ga$_2$O$_3$, to enhance breakdown characteristics as well as present a comparative analysis of MIS and FP-MS structures. Performance of the two modified diodes has been compared in terms of parameters such as $V_{BR}$, specific on-resistance ($R_{on,sp}$), and Schottky barrier height (SBH). BZN integration enhances surface E-field redistribution in both the modified diodes compared to the control metal-semiconductor (MS) device resulting in nearly two-fold improvement in $V_{BR}$, with minimal impact on $R_{on,sp}$. The enhancement in Schottky barrier height (SBH) by approximately 0.14 eV for MIS and 0.28 eV for FP-MS devices also contributes to the improvement in $V_{BR}$. The devices display excellent current-voltage characteristics with close-to-unity ideality factors and high on/off current ratio ($I_{on}/I_{off}$ $>$ 10$^{10}$). Specifically, the absolute $V_{BR}$ value increases from 300 V for the MS SBD to 600 V for both FP-MS and MIS SBDs, also the best reported value for MIS diodes. However, due to insertion of an insulator layer at the metal-semiconductor junction, $R_{on,sp}$ is slightly higher for MIS SBDs i.e. 2.35 m$\Omega$.cm$^2$ compared to 1.98 m$\Omega$.cm$^2$ for FP-MS. These parameters show the effectiveness of BZN (k $\sim$210) in enhancing the breakdown electric field in $\beta$-Ga$_2$O$_3$ approaching its breakdown field limit under reverse bias. Additionally, the FP-MS SBD architecture displays greater promise for switching applications with smaller conduction losses when compared to MIS SBDs. The dielectric constant of BZN was determined to be $\sim$210 from metal-BZN-metal (MIM) capacitance-voltage (C-V) profiles. Its optical bandgap, $\sim$3.3 eV, was obtained from the Tauc plot, and the amorphous nature of the film was confirmed using grazing incidence x-ray diffraction (GIXRD). 

\section*{\textbf{Experimental}}

For comparison, three types of vertical SBDs i.e. Pt/$\beta$-Ga$_2$O$_3$/Ti (MS), Pt/BZN/$\beta$-Ga$_2$O$_3$/Ti (MIS and FP-MS)  were fabricated on three 5 x 5 mm$^2$, 7 $\mu$m thick, 4 x 10$^{16}$ cm$^{-3}$ doped epi-layer deposited $\beta$-Ga$_2$O$_3$ conducting substrates diced from a 2" wafer obtained from Novel Crystal Technology Inc. The substrates were first cleaned with a series of ultrasonicated baths in methanol, acetone, and deionized (DI) water to eliminate surface contaminants. Following this, a piranha solution treatment was carried out for 5 minutes to ensure removal of any remaining organic residues from the surface, followed by a final rinse with DI water \cite{biswas2020, Biswas2019, PSharma_ICEE2022, sharma2024monolithic}. 
Ti/Au (30/70 nm) backside substrate contacts were deposited via DC sputtering followed by rapid thermal annealing (RTA) at 470 $\degree$C for 1 minute in an N$_2$ gas ambient for improved Ohmic contacts.
Thick ($\sim$280 nm for FP-MS) and thin (20 nm for MIS) BZN layers were deposited on two samples separately via RF sputtering. The deposition process was performed with RF power of 150 W at a substrate temperature of 350 $\degree$C maintaining an Ar:O$_2$ gas flow rate ratio of 35:15 sccm. For FP-MS SBDs, the 280 nm film was patterned using HF (2\%) etch. Circular top Schottky contacts, 200 μm in diameter, were defined on all three samples (MS, FP-MS and MIS SBDs shown in Fig. \ref{fig:fig0}) using optical lithography, followed by Pt/Au (30/70 nm) deposition using DC sputtering and lift-off. The device cross-sections are depicted in Fig. \ref{fig:fig0}(a), (b) and (c).

The high-k BZN dielectric film was also deposited onto a quartz glass substrate for optical characterization to determine its bandgap. Morphological properties of the film were analyzed using x-ray photoelectron spectroscopy (XPS) and x-ray diffractometry (XRD). Metal-insulator-metal (MIM) capacitors with a Pt/BZN ($\sim$20 and 280 nm)/Pt stack were also fabricated on an Si/SiO$_2$ substrate, as depicted in the inset of Fig. \ref{fig:fig0}(d).

    \label{fig:fig1_3}

\section*{\textbf{Results and Discussion}}
Pyrochlore BZN has been reported in literature as a low-loss, high-permittivity dielectric material, and the dielectric constant is insensitive to variations in film thickness. The dielectric constant of  BZN was determined by measuring $C-V$ profiles of the fabricated MIM capacitors as shown in Fig. \ref{fig:fig0}(d). Inset shows the device cross-section. Extracted dielectric constant value of 210 is close to values reported\cite{lu2003low, lee2008dielectric, PSedtm2024} for films deposited using RF sputtering. The 3.3 eV optical bandgap of BZN, determined using a Tauc plot (refer supplementary section 1), is also in close agreement with the reported value\cite{qasrawi2014electrical}. GIXRD (refer supplementary section 1) confirmed amorphous nature of the deposited film. Analysis of XPS spectra (supplementary section 2) collected from the BZN film surface revealed a chemical composition of Bi$_{0.82}$Zn$_1$Nb$_{0.94}$O$_7$. This composition matches well with that of the sputter target (Bi$_{1.5}$Zn$_1$Nb$_{1.5}$O$_7$). 

\begin{figure}[h]
    \centering 
    \includegraphics[width=1\linewidth]{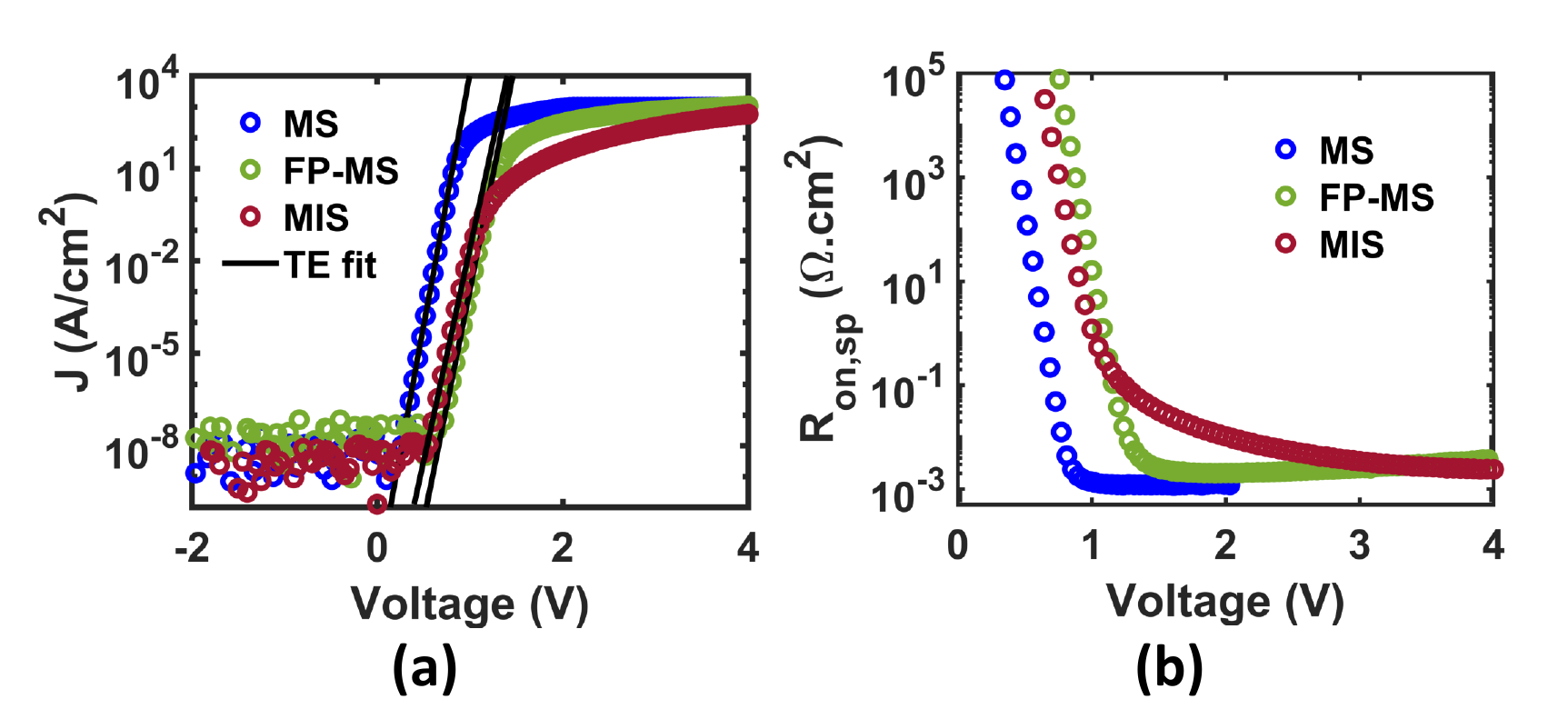}
        \caption{(a) Forward bias $J-V$ characteristics measured at room temperature, and, (b) specific on-resistance, $R_{on,sp}$ of the fabricated MS, FP-MS and MIS SBDs.}
    \label{fig:fig2}
\end{figure}

The MS, FP-MS and MIS SBDs fabricated on $\beta$-Ga$_2$O$_3$ were electrically characterized using a Keysight B1505 semiconductor parameter analyzer. The forward biased current-voltage ($J-V$) characteristics of the three devices measured at room temperature, shown in Fig. \ref{fig:fig2}(b), were analyzed under the assumption that thermionic emission is the dominant  current transport mechanism \cite{sze2021physics},
\begin{equation}
J=J_s(e^{qV/\eta kT} -1),
\label{eq:1}
\end{equation}
where,
\begin{equation}
J_s = A^{**}T^2 e^{-q\phi_B/kT}
\end{equation}
Here, $J_s$ is the reverse saturation current density, $q$ is electron charge, $V$ is applied bias voltage, $\eta$ is the ideality factor, $k$ is Boltzmann constant, $A^{**}$ is the Richardson constant, $T$ is the temperature, $\phi_B$ is the SBH. We have assumed $A^{**}$= 41.1 A/cm$^{2}$K$^{2}$, calculated using an electron effective mass of 0.34 m$_e$ \cite{PeartonAPR20182} for $\beta$-Ga$_2$O$_3$, to extract the SBH from $J-V$ data measured at room temperature.
The MS, FP-MS, and MIS devices show ideality factors ($\eta$) of 1.02, 1.12, and 1.21, respectively, and an $I_{on}$/$I_{off}$ ratio of over 10$^{10}$, indicating good device quality. The three types of devices exhibit low reverse leakage current densities in the nA/cm$^2$ range and turn-on voltages of 0.70 V for MS, 1.25 V for MIS, and 1.20 V for FP-MS as shown in Fig. \ref{fig:fig2}. 
Incorporation of the BZN high-k dielectric film had minimal impact on the $R_{on,sp}$ with values of 1.16, 1.98, and 2.35 mΩ.cm$^2$ for the MS, FP-MS and MIS  SBDs, respectively. The average values for key Schottky diode parameters are tabulated in Table \ref{tab:tab1} for the three type of devices. 

\begin{figure}[h]
    \centering 
    \includegraphics[width=1\linewidth]{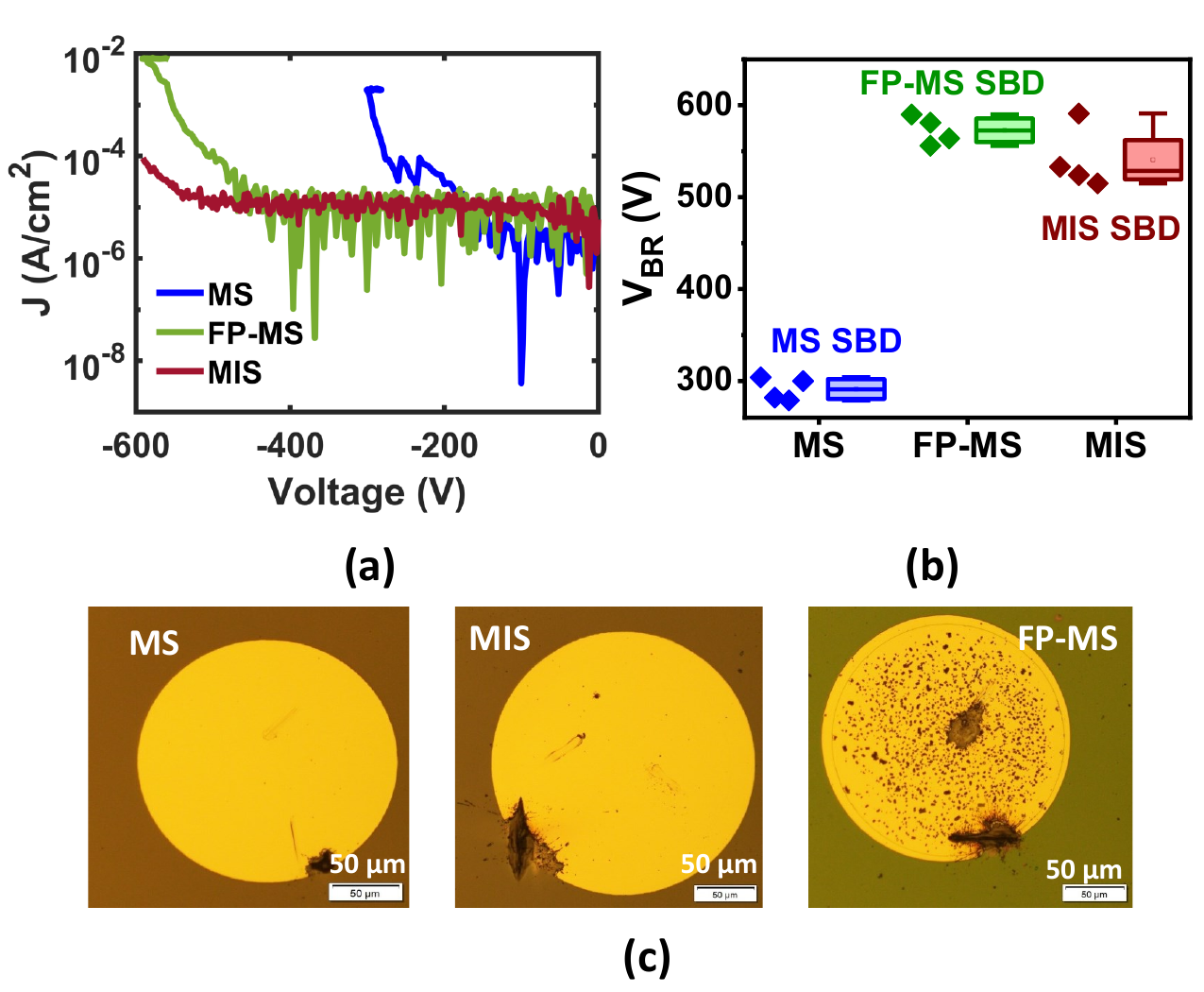}
        \caption{(a) Reverse bias $J-V$ characteristics measured at room temperature, (b)
$V_{BR}$ distribution for the three types of 
 devices fabricated on three 5 x 5 mm$^2$ substrates taken from the 2" $\beta$-Ga$_2$O$_3$ wafer, and (c) post catastrophic breakdown images of fabricated devices.}
    \label{fig:fig3}
\end{figure}

\begin{table*}
\caption{\label{tab:tab1}Electrical parameters ($\eta$, R$_{on,sp}$, SBH and $V_{BR}$) obtained from $J-V$ characteristics measured for fabricated MS, FP-Ms and MIS SBDs.  }


\begin{tabular}{lccccc}
\hline
 Device&$\eta$ &R$_{on,sp}$ & $\phi_B$ & $V_{ON}$&\textit{V$_{BR}$} \\
 & & (m$\Omega$.cm$^2$)& (eV)& (V)& (V) \\
 
\hline
MS& 1.02 ± 0.02& 1.16 ± 0.04& 1.16 ± 0.01&  0.70&291 ± 11\\
FP-MS& 1.12 ± 0.05& 1.98 ± 0.07& 1.44 ± 0.03& 1.20&573 ± 13\\
MIS& 1.21 ± 0.04& 2.35 ± 0.06& 1.30 ± 0.04&  1.25&540 ± 30\\
 \hline
\end{tabular}
\end{table*}

\begin{figure*}[h]
    \centering 
    \includegraphics[width=1\linewidth]{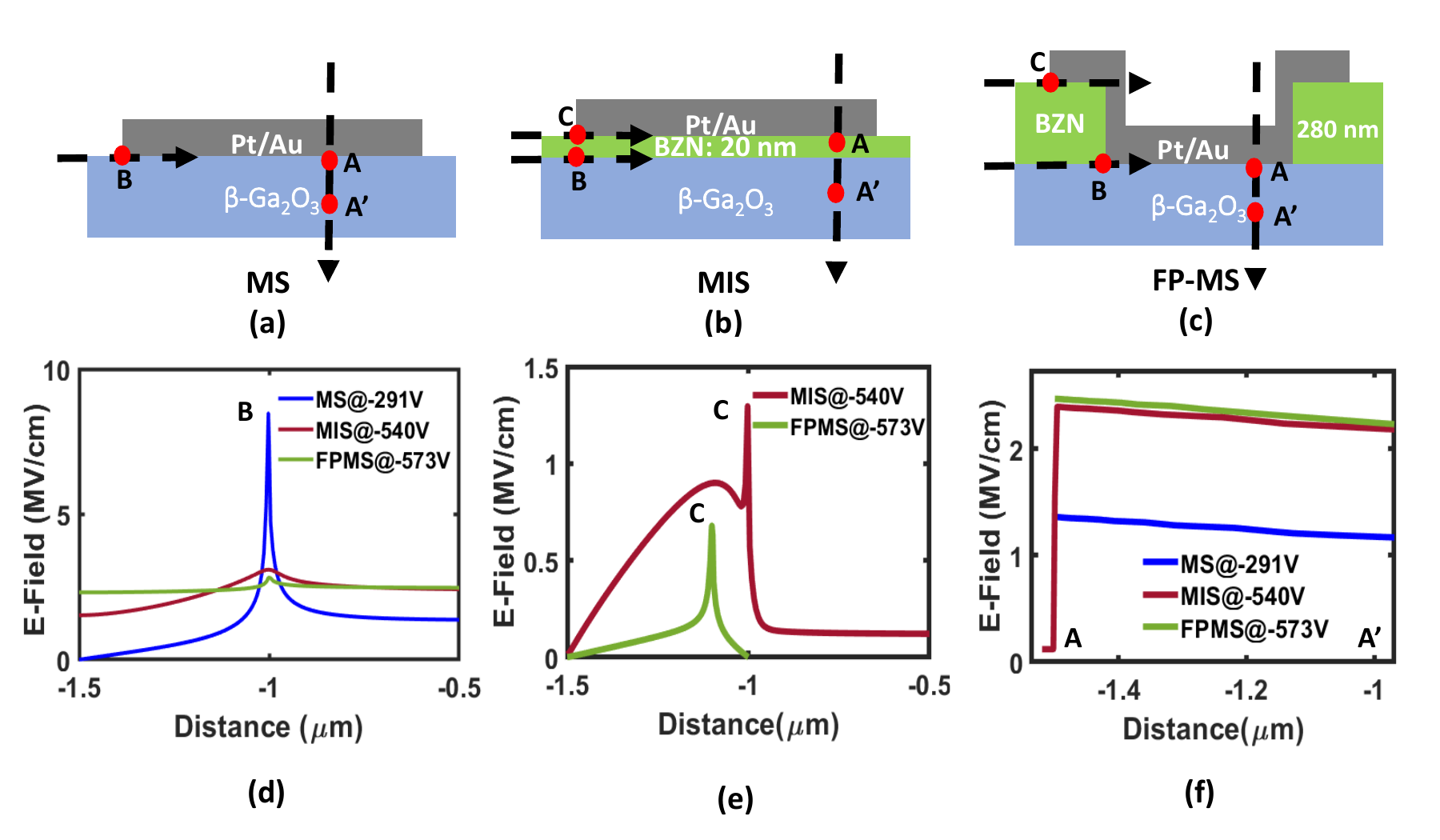}
        \caption{Cross-sections of (a) MS, (b) MIS, (c) FP-MS diodes  highlighting critical points for field crowding. Horizontal field distributions through points (d) B and (e) C illustrate reduced corner electric fields at breakdown voltages in the two BZN-based SBDs. Additionally, the field profile across (f) A and A' indicates reduced electric field at the metal/$\beta$-Ga$_2$O$_3$ interface and increased breakdown electric fields in the substrate for MIS and FP-MS SBDs at $V_{BR}$. }
    \label{fig:fig_tcad}
\end{figure*}

The inclusion of BZN increased the SBH from 1.16 eV for the MS to 1.44 eV for the FP-MS and 1.30 eV for MIS SBDs. The improved SBH for FP-MS is likely due to exposure to a 2$\%$ HF solution during BZN etch, which resulted in surface passivation with fluorine and subsequent Fermi level unpinning \cite{psharmaAPL2024, Konishi2017}. In the case of MIS, the SBH increase is likely due to Fermi level unpinning caused by the insertion of a dielectric at the metal-semiconductor interface \cite{FLP2023reduction, bhattacharyya2020schottky}. The enhanced SBH and high dielectric constant of BZN (improved field distribution) together lead to a remarkable improvement in $V_{BR}$ as shown in Fig. \ref{fig:fig3}(b). It increased by nearly 2x from 300 V for the MS SBD to $\sim$600 V for the MIS and FP-MS devices.  The measured $V_{BR}$ for the MS SBD is consistent with values reported for devices fabricated on substrates with similar doping level and drift layer thickness.  \cite{dhara2022beta, he2021over}. In case of FP-MS, high-k incorporation enhances uniformity of the field distribution and as a result $V_{BR}$ is higher. The MIS diode is additionally helped by the fact that incorporating BZN effectively reduces the electric field at the metal-dielectric interface by leveraging dielectric discontinuity ($\epsilon_{BZN}$/$\epsilon_{Ga_2O_3}$ $\sim$21). TCAD simulations of the three device architectures as shown in Fig. \ref{fig:fig_tcad}(a), (b), and (c) were carried out to extract E-field profiles at measured breakdown voltages. 
The E-field profiles shown in Fig. \ref{fig:fig_tcad}(d) confirm reduction in the electric field at the semiconductor surface/interface, particularly near the metal contact edges (marked as point B), achieved by the insertion of the high-k BZN layer. Used either as a field plate or as a thin IL, it effectively reduces the surface field. It is important to note that the field reduction in the MIS device is less pronounced than in the FP-MS case due to the difference in BZN layer thickness between the two configurations. Thicker dielectric layers are more effective at redistributing edge or corner fields, but they can increase specific on-resistance ($R_{on,sp}$) when used in MIS devices. Therefore, thinner layers are preferred for MIS applications. Similarly, the field at the metal edges (marked as point C) is also reduced when the metal edge is not in direct contact with the semiconductor but on top of the BZN film, as illustrated in Fig. \ref{fig:fig_tcad}(e). 
Beyond the edge electric field reduction, the MIS device additionally benefits from a reduction in the metal/BZN interface electric field in the bulk of the device as shown by the E-field profiles along A-A' in Fig. \ref{fig:fig_tcad}(f). It can also be seen from these profiles that the field reduction due to BZN allows a higher $V_{BR}$ for MIS and FP-MS diodes resulting in a higher $\beta$-Ga$_2$O$_3$ breakdown field, compared to MS devices, reaching a maximum value of 2.4 MV/cm for the FP-MS SBDs.


Comparison of the characteristics of the FP-MS and MIS devices indicates that FP-MS devices perform better as there is less impact on $R_{on,sp}$ while the $V_{BR}$ gained is also slightly more than for MIS devices based on the average value of four devices each as shown in Fig.\ref{fig:fig3}(b). Highest $V_{BR}$ measured for the two BZN-based devices is 590 V. In MIS, the high-k dielectric helps in reducing the electric field at the metal-dielectric junction, in the bulk as well as at the edge, allowing an increase in the field in the bulk of $\beta$-Ga$_2$O$_3$ which ultimately increases the $V_{BR}$. Three factors contribute to the enhancement of $V_{BR}$: the high-k BZN as an interlayer (IL), its presence at the edges, and the improved Schottky barrier height (SBH). In the FP-MS device, only two of these factors contribute: the enhanced SBH and edge field reduction with high-k BZN. It is important to note, however, that the edge electric field reduction is better for the FP-MS device due to thicker BZN, that results in higher $V_{BR}$ as compared to MIS devices. This has been confirmed by extracting field profiles from TCAD simulations of the three device architectures as shown in Fig. \ref{fig:fig_tcad}. By fabricating FP-MS diodes, the breakdown problem at device edge due to field crowding is largely addressed but the breakdown still seems to initiate at the Schottky contact edge as shown in Fig. \ref{fig:fig3}(c). Hence, edge field termination appears to be a more critical challenge than interface optimization when it comes to improving the breakdown fields of $\beta$-Ga$_2$O$_3$ devices.

\begin{figure}[ht]
    \centering
    \includegraphics[width=0.75\linewidth]{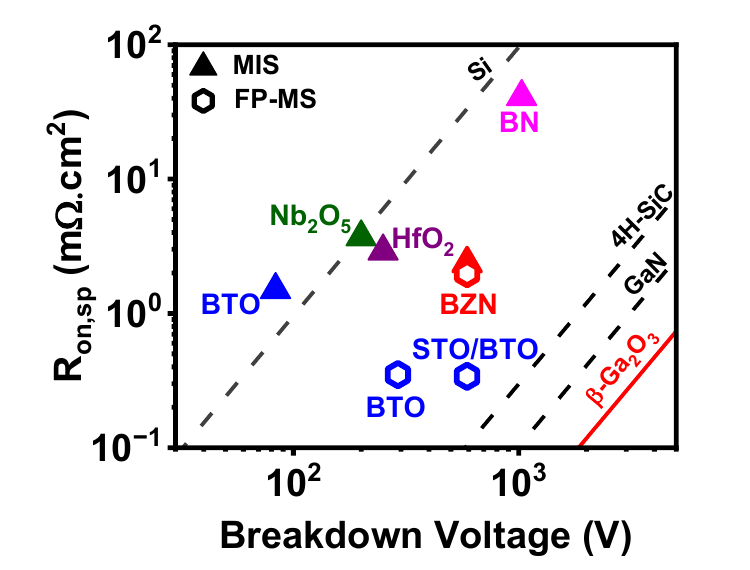}
    \caption{Benchmarking of R$_{on,sp}$ and \textit{V$_{BR}$} values from this work with state-of-the-art devices featuring various Schottky diode designs with high-k dielectric integration.}
    \label{fig:fig4}
\end{figure}

Benchmarking the $V_{BR}$ and $R_{on,sp}$ data, as shown in Fig. \ref{fig:fig4}, from this experiment against reported MIS \cite{xia2019metal, tiwari2021nb2o5, sasaki2017first, xu2023_BN} and high-k FP-MS \cite{roy2021high} SBDs shows that the high-k BZN MIS fabricated in this study outperforms previously reported MIS devices. While it is established, both in this study and in BTO-based research, that high-k FP-MS devices generally perform better than high-k MIS devices, further improvement in breakdown performance will require combining high-k dielectric field plates with additional edge field termination techniques to further reduce peak electric fields at the device edges. A combination of MIS and FP-MS structures, such as an FP-MIS device, with additional edge termination techniques appears to be a promising solution for further enhancing device performance by effectively reducing peak electric fields at the edges along with addressing the interface fermi level pinning with the same high-k dielectric material. 

\section*{\textbf{Conclusion}}
In conclusion, this study has demonstrated the fabrication and characterization of $\beta$-Ga$_2$O$_3$ SBDs incorporating the high-k dielectric BZN deposited at 350 $\degree$C by RF sputtering. Integrating BZN led to a significant improvement in the SBH, increasing it from 1.16 eV to 1.30 and 1.44 eV in MIS and FP-MS SBDs. This enhancement, along with improved edge electric field distribution, translates into a remarkable two-fold improvement in the $V_{BR}$, increasing it from 300 V to 600 V while maintaining a low R$_{on,sp}$. The two modified devices exhibit excellent electrical $J-V$ characteristics, with close to unity ideality factor and $I_{on}/I_{off}>10^{10}$. We report the largest enhancement in $V_{BR}$ of $\beta$-Ga$_2$O$_3$ SBDs using a high-k dielectric interlayer at the metal-semiconductor interface, while keeping the on-resistance value in check. Additionally, the FP-MS structure has shown superior performance compared to the MIS diode.
These findings underscore the potential of high-k dielectric materials in enhancing the performance of $\beta$-Ga$_2$O$_3$ Schottky diodes for high-power applications. 

\section*{Supplementary Material}
The supplementary material includes a table comparing the deposition temperatures of various interlayer dielectrics reported on $\beta$-Ga$_2$O$_3$ with that of BZN (section 1), a Tauc plot showing optical bandgap extraction and GIXRD data of the BZN film (section 2), BZN XPS data (section 3), MIS turn-on voltage benchmarking and additional experimental details.

\section*{\textbf{Acknowledgment}}
The authors thank Dr.  KP Sreejith for UV-vis spectrum measurements. The authors acknowledge the Indian Institute of Technology Bombay Nano-fabrication Facility (IITBNF) for the usage of its device fabrication and characterization facilities and support from the Ministry of Electronics and Information Technology (through project 5(1)/2017-NANO) and Department of Science and Technology (through project DST/NM/NNetRA/2018(G)-IIT-B), Government of India, for funding this work. Pooja Sharma thanks University Grants Commission (UGC) for the doctoral fellowship. 

\section*{Data Availability}
The data supporting this study's findings are included within the article [and its supplementary material]. 


\bibliography{ref2}

\providecommand{\latin}[1]{#1}
\makeatletter
\providecommand{\doi}
  {\begingroup\let\do\@makeother\dospecials
  \catcode`\{=1 \catcode`\}=2 \doi@aux}
\providecommand{\doi@aux}[1]{\endgroup\texttt{#1}}
\makeatother
\providecommand*\mcitethebibliography{\thebibliography}
\csname @ifundefined\endcsname{endmcitethebibliography}  {\let\endmcitethebibliography\endthebibliography}{}
\begin{mcitethebibliography}{34}
\providecommand*\natexlab[1]{#1}
\providecommand*\mciteSetBstSublistMode[1]{}
\providecommand*\mciteSetBstMaxWidthForm[2]{}
\providecommand*\mciteBstWouldAddEndPuncttrue
  {\def\EndOfBibitem{\unskip.}}
\providecommand*\mciteBstWouldAddEndPunctfalse
  {\let\EndOfBibitem\relax}
\providecommand*\mciteSetBstMidEndSepPunct[3]{}
\providecommand*\mciteSetBstSublistLabelBeginEnd[3]{}
\providecommand*\EndOfBibitem{}
\mciteSetBstSublistMode{f}
\mciteSetBstMaxWidthForm{subitem}{(\alph{mcitesubitemcount})}
\mciteSetBstSublistLabelBeginEnd
  {\mcitemaxwidthsubitemform\space}
  {\relax}
  {\relax}

\bibitem[Sharma and Lodha(2024)Sharma, and Lodha]{psharmaAPL2024}
Sharma,~P.; Lodha,~S. { $\beta$-Ga$_2$O$_3$ Schottky barrier height improvement using Ar/O$_2$ plasma and HF surface treatments}. \emph{Applied Physics Letters} \textbf{2024}, \emph{124}, 072106\relax
\mciteBstWouldAddEndPuncttrue
\mciteSetBstMidEndSepPunct{\mcitedefaultmidpunct}
{\mcitedefaultendpunct}{\mcitedefaultseppunct}\relax
\EndOfBibitem
\bibitem[Yang \latin{et~al.}(2018)Yang, Sparks, Ren, Pearton, and Tadjer]{JYang2018}
Yang,~J.; Sparks,~Z.; Ren,~F.; Pearton,~S.~J.; Tadjer,~M. Effect of surface treatments on electrical properties of $\beta$-Ga$_2$O$_3$. \emph{Journal of Vacuum Science \& Technology B} \textbf{2018}, \emph{36}\relax
\mciteBstWouldAddEndPuncttrue
\mciteSetBstMidEndSepPunct{\mcitedefaultmidpunct}
{\mcitedefaultendpunct}{\mcitedefaultseppunct}\relax
\EndOfBibitem
\bibitem[Bhattacharyya \latin{et~al.}(2020)Bhattacharyya, Ranga, Saleh, Roy, Scarpulla, Lynn, and Krishnamoorthy]{bhattacharyya2020schottky}
Bhattacharyya,~A.; Ranga,~P.; Saleh,~M.; Roy,~S.; Scarpulla,~M.~A.; Lynn,~K.~G.; Krishnamoorthy,~S. Schottky barrier height engineering in $\beta$-Ga$_2$O$_3$ using SiO$_2$ interlayer dielectric. \emph{IEEE Journal of the Electron Devices Society} \textbf{2020}, \emph{8}, 286--294\relax
\mciteBstWouldAddEndPuncttrue
\mciteSetBstMidEndSepPunct{\mcitedefaultmidpunct}
{\mcitedefaultendpunct}{\mcitedefaultseppunct}\relax
\EndOfBibitem
\bibitem[Prajapati and Lodha(2024)Prajapati, and Lodha]{Prajapati2024}
Prajapati,~P.; Lodha,~S. Barrier height enhancement in $\beta$-Ga$_2$O$_3$ Schottky diodes using an oxygen-rich ultra-thin AlO$_x$ interfacial layer. \emph{Applied Physics Letters} \textbf{2024}, \emph{125}\relax
\mciteBstWouldAddEndPuncttrue
\mciteSetBstMidEndSepPunct{\mcitedefaultmidpunct}
{\mcitedefaultendpunct}{\mcitedefaultseppunct}\relax
\EndOfBibitem
\bibitem[Xu \latin{et~al.}(2023)Xu, Biswas, Li, He, Luo, Mei, Zhou, Chang, Puthirath, Vajtai, \latin{et~al.} others]{xu2023_BN}
Xu,~M.; Biswas,~A.; Li,~T.; He,~Z.; Luo,~S.; Mei,~Z.; Zhou,~J.; Chang,~C.; Puthirath,~A.~B.; Vajtai,~R.; others Vertical $\beta$-Ga$_2$O$_3$ metal--insulator--semiconductor diodes with an ultrathin boron nitride interlayer. \emph{Applied Physics Letters} \textbf{2023}, \emph{123}\relax
\mciteBstWouldAddEndPuncttrue
\mciteSetBstMidEndSepPunct{\mcitedefaultmidpunct}
{\mcitedefaultendpunct}{\mcitedefaultseppunct}\relax
\EndOfBibitem
\bibitem[Lu \latin{et~al.}(2022)Lu, Xu, Deng, Liao, Luo, Pei, Chen, Lv, and Wang]{Lu2022_NiO}
Lu,~X.; Xu,~T.; Deng,~Y.; Liao,~C.; Luo,~H.; Pei,~Y.; Chen,~Z.; Lv,~Y.; Wang,~G. Performance-enhanced NiO/$\beta$-Ga$_2$O$_3$ heterojunction diodes fabricated on an etched $\beta$-Ga$_2$O$_3$ surface. \emph{Applied Surface Science} \textbf{2022}, \emph{597}, 153587\relax
\mciteBstWouldAddEndPuncttrue
\mciteSetBstMidEndSepPunct{\mcitedefaultmidpunct}
{\mcitedefaultendpunct}{\mcitedefaultseppunct}\relax
\EndOfBibitem
\bibitem[Hao \latin{et~al.}(2022)Hao, He, Zhou, Zhao, Xu, and Long]{Hao2022_NiO}
Hao,~W.; He,~Q.; Zhou,~X.; Zhao,~X.; Xu,~G.; Long,~S. 2.6 kV NiO/Ga$_2$O$_3$ heterojunction diode with superior high-temperature voltage blocking capability. \emph{2022 IEEE 34th International Symposium on Power Semiconductor Devices and ICs (ISPSD)} \textbf{2022}, 105--108\relax
\mciteBstWouldAddEndPuncttrue
\mciteSetBstMidEndSepPunct{\mcitedefaultmidpunct}
{\mcitedefaultendpunct}{\mcitedefaultseppunct}\relax
\EndOfBibitem
\bibitem[Watahiki \latin{et~al.}(2017)Watahiki, Yuda, Furukawa, Yamamuka, Takiguchi, and Miyajima]{Watahiki2017_Cu2O}
Watahiki,~T.; Yuda,~Y.; Furukawa,~A.; Yamamuka,~M.; Takiguchi,~Y.; Miyajima,~S. Heterojunction p-Cu2O/n-Ga$_2$O$_3$ diode with high breakdown voltage. \emph{Applied Physics Letters} \textbf{2017}, \emph{111}\relax
\mciteBstWouldAddEndPuncttrue
\mciteSetBstMidEndSepPunct{\mcitedefaultmidpunct}
{\mcitedefaultendpunct}{\mcitedefaultseppunct}\relax
\EndOfBibitem
\bibitem[Sun \latin{et~al.}(2018)Sun, Liao, Chen, Li, Lin, and Song]{sun2018_NiO_mobility}
Sun,~H.; Liao,~M.-H.; Chen,~S.-C.; Li,~Z.-Y.; Lin,~P.-C.; Song,~S.-M. Synthesis and characterization of n-type NiO: Al thin films for fabrication of pn NiO homojunctions. \emph{Journal of Physics D: Applied Physics} \textbf{2018}, \emph{51}, 105109\relax
\mciteBstWouldAddEndPuncttrue
\mciteSetBstMidEndSepPunct{\mcitedefaultmidpunct}
{\mcitedefaultendpunct}{\mcitedefaultseppunct}\relax
\EndOfBibitem
\bibitem[Alidoust and Carter(2015)Alidoust, and Carter]{alidoust2015_NiO_mobility}
Alidoust,~N.; Carter,~E.~A. First-principles assessment of hole transport in pure and Li-doped NiO. \emph{Physical Chemistry Chemical Physics} \textbf{2015}, \emph{17}, 18098--18110\relax
\mciteBstWouldAddEndPuncttrue
\mciteSetBstMidEndSepPunct{\mcitedefaultmidpunct}
{\mcitedefaultendpunct}{\mcitedefaultseppunct}\relax
\EndOfBibitem
\bibitem[Tiwari \latin{et~al.}(2021)Tiwari, Biswas, Joishi, and Lodha]{tiwari2021nb2o5}
Tiwari,~P.; Biswas,~J.; Joishi,~C.; Lodha,~S. Nb$_2$O$_5$ high-k dielectric enabled electric field engineering of $\beta$-Ga$_2$O$_3$ metal-insulator-semiconductor (MIS) diode. \emph{Journal of Applied Physics} \textbf{2021}, \emph{130}\relax
\mciteBstWouldAddEndPuncttrue
\mciteSetBstMidEndSepPunct{\mcitedefaultmidpunct}
{\mcitedefaultendpunct}{\mcitedefaultseppunct}\relax
\EndOfBibitem
\bibitem[Roy \latin{et~al.}(2021)Roy, Bhattacharyya, Ranga, Splawn, Leach, and Krishnamoorthy]{roy2021high}
Roy,~S.; Bhattacharyya,~A.; Ranga,~P.; Splawn,~H.; Leach,~J.; Krishnamoorthy,~S. High-k oxide field-plated vertical $(001)$ $\beta$-Ga$_2$O$_3$ Schottky barrier diode with Baliga’s figure of merit over 1 GW/cm$^2$. \emph{IEEE Electron Device Letters} \textbf{2021}, \emph{42}, 1140--1143\relax
\mciteBstWouldAddEndPuncttrue
\mciteSetBstMidEndSepPunct{\mcitedefaultmidpunct}
{\mcitedefaultendpunct}{\mcitedefaultseppunct}\relax
\EndOfBibitem
\bibitem[Xia \latin{et~al.}(2019)Xia, Chandrasekar, Moore, Wang, Lee, McGlone, Kalarickal, Arehart, Ringel, Yang, \latin{et~al.} others]{xia2019metal}
Xia,~Z.; Chandrasekar,~H.; Moore,~W.; Wang,~C.; Lee,~A.~J.; McGlone,~J.; Kalarickal,~N.~K.; Arehart,~A.; Ringel,~S.; Yang,~F.; others Metal/BaTiO$_3$/$\beta$-Ga$_2$O$_3$ dielectric heterojunction diode with 5.7 MV/cm breakdown field. \emph{Applied Physics Letters} \textbf{2019}, \emph{115}\relax
\mciteBstWouldAddEndPuncttrue
\mciteSetBstMidEndSepPunct{\mcitedefaultmidpunct}
{\mcitedefaultendpunct}{\mcitedefaultseppunct}\relax
\EndOfBibitem
\bibitem[Roy \latin{et~al.}(2023)Roy, Kostroun, Cooke, Liu, Bhattacharyya, Peterson, Sensale-Rodriguez, and Krishnamoorthy]{roy2023BTO}
Roy,~S.; Kostroun,~B.; Cooke,~J.; Liu,~Y.; Bhattacharyya,~A.; Peterson,~C.; Sensale-Rodriguez,~B.; Krishnamoorthy,~S. Ultra-low reverse leakage in large area kilo-volt class $\beta$-Ga$_2$O$_3$ trench Schottky barrier diode with high-k dielectric RESURF. \emph{Applied Physics Letters} \textbf{2023}, \emph{123}\relax
\mciteBstWouldAddEndPuncttrue
\mciteSetBstMidEndSepPunct{\mcitedefaultmidpunct}
{\mcitedefaultendpunct}{\mcitedefaultseppunct}\relax
\EndOfBibitem
\bibitem[Roy \latin{et~al.}(2023)Roy, Bhattacharyya, Peterson, and Krishnamoorthy]{roy2023_BTO2}
Roy,~S.; Bhattacharyya,~A.; Peterson,~C.; Krishnamoorthy,~S. 2.1 kV (001)-$\beta$-Ga$_2$O$_3$ vertical Schottky barrier diode with high-k oxide field plate. \emph{Applied Physics Letters} \textbf{2023}, \emph{122}\relax
\mciteBstWouldAddEndPuncttrue
\mciteSetBstMidEndSepPunct{\mcitedefaultmidpunct}
{\mcitedefaultendpunct}{\mcitedefaultseppunct}\relax
\EndOfBibitem
\bibitem[Rahman \latin{et~al.}(2021)Rahman, Kalarickal, Lee, Razzak, and Rajan]{rahman2021_BTO}
Rahman,~M.~W.; Kalarickal,~N.~K.; Lee,~H.; Razzak,~T.; Rajan,~S. Integration of high permittivity BaTiO$_3$ with AlGaN/GaN for near-theoretical breakdown field kV-class transistors. \emph{Applied Physics Letters} \textbf{2021}, \emph{119}\relax
\mciteBstWouldAddEndPuncttrue
\mciteSetBstMidEndSepPunct{\mcitedefaultmidpunct}
{\mcitedefaultendpunct}{\mcitedefaultseppunct}\relax
\EndOfBibitem
\bibitem[Razzak \latin{et~al.}(2020)Razzak, Chandrasekar, Hussain, Lee, Mamun, Xue, Xia, Sohel, Rahman, Bajaj, \latin{et~al.} others]{razzak2020_BTO}
Razzak,~T.; Chandrasekar,~H.; Hussain,~K.; Lee,~C.~H.; Mamun,~A.; Xue,~H.; Xia,~Z.; Sohel,~S.~H.; Rahman,~M.~W.; Bajaj,~S.; others BaTiO$_3$/Al$_{0. 58}$Ga$_{0. 42}$N lateral heterojunction diodes with breakdown field exceeding 8 MV/cm. \emph{Applied Physics Letters} \textbf{2020}, \emph{116}\relax
\mciteBstWouldAddEndPuncttrue
\mciteSetBstMidEndSepPunct{\mcitedefaultmidpunct}
{\mcitedefaultendpunct}{\mcitedefaultseppunct}\relax
\EndOfBibitem
\bibitem[Rahman \latin{et~al.}(2021)Rahman, Chandrasekar, Razzak, Lee, and Rajan]{rahman2021_BTO2}
Rahman,~M.~W.; Chandrasekar,~H.; Razzak,~T.; Lee,~H.; Rajan,~S. Hybrid BaTiO$_3$/SiN$_x$/AlGaN/GaN lateral Schottky barrier diodes with low turn-on and high breakdown performance. \emph{Applied Physics Letters} \textbf{2021}, \emph{119}\relax
\mciteBstWouldAddEndPuncttrue
\mciteSetBstMidEndSepPunct{\mcitedefaultmidpunct}
{\mcitedefaultendpunct}{\mcitedefaultseppunct}\relax
\EndOfBibitem
\bibitem[Biswas \latin{et~al.}(2020)Biswas, Joishi, Biswas, Tiwari, and Lodha]{biswas2020}
Biswas,~D.; Joishi,~C.; Biswas,~J.; Tiwari,~P.; Lodha,~S. {Charge trap layer enabled positive tunable V$_{FB}$ in $\beta$-Ga$_2$O$_3$ gate stacks for enhancement mode transistors}. \emph{Applied Physics Letters} \textbf{2020}, \emph{117}, 172101\relax
\mciteBstWouldAddEndPuncttrue
\mciteSetBstMidEndSepPunct{\mcitedefaultmidpunct}
{\mcitedefaultendpunct}{\mcitedefaultseppunct}\relax
\EndOfBibitem
\bibitem[Biswas \latin{et~al.}(2019)Biswas, Joishi, Biswas, Thakar, Rajan, and Lodha]{Biswas2019}
Biswas,~D.; Joishi,~C.; Biswas,~J.; Thakar,~K.; Rajan,~S.; Lodha,~S. {Enhanced n-type $\beta$-Ga$_2$O$_3$ ($\bar{2}$01) gate stack performance using Al$_2$O$_3$/SiO$_2$ bi-layer dielectric}. \emph{Applied Physics Letters} \textbf{2019}, \emph{114}, 212106\relax
\mciteBstWouldAddEndPuncttrue
\mciteSetBstMidEndSepPunct{\mcitedefaultmidpunct}
{\mcitedefaultendpunct}{\mcitedefaultseppunct}\relax
\EndOfBibitem
\bibitem[Sharma and Lodha(2022)Sharma, and Lodha]{PSharma_ICEE2022}
Sharma,~P.; Lodha,~S. Effect of HF Surface Treatment on Electrical Properties of $\beta$-Ga$_2$O$_3$ Schottky Barrier Diodes. 2022 IEEE International Conference on Emerging Electronics (ICEE). 2022; pp 1--3\relax
\mciteBstWouldAddEndPuncttrue
\mciteSetBstMidEndSepPunct{\mcitedefaultmidpunct}
{\mcitedefaultendpunct}{\mcitedefaultseppunct}\relax
\EndOfBibitem
\bibitem[Sharma and Lodha(2024)Sharma, and Lodha]{sharma2024monolithic}
Sharma,~P.; Lodha,~S. Monolithic $\beta$-Ga$_2$O$_3$ Bidirectional MOSFET. \emph{arXiv preprint arXiv:2407.17263} \textbf{2024}, \relax
\mciteBstWouldAddEndPunctfalse
\mciteSetBstMidEndSepPunct{\mcitedefaultmidpunct}
{}{\mcitedefaultseppunct}\relax
\EndOfBibitem
\bibitem[Lu and Stemmer(2003)Lu, and Stemmer]{lu2003low}
Lu,~J.; Stemmer,~S. Low-loss, tunable bismuth zinc niobate films deposited by rf magnetron sputtering. \emph{Applied Physics Letters} \textbf{2003}, \emph{83}, 2411--2413\relax
\mciteBstWouldAddEndPuncttrue
\mciteSetBstMidEndSepPunct{\mcitedefaultmidpunct}
{\mcitedefaultendpunct}{\mcitedefaultseppunct}\relax
\EndOfBibitem
\bibitem[Lee \latin{et~al.}(2008)Lee, Lee, Song, Lee, and Chung]{lee2008dielectric}
Lee,~S.~E.; Lee,~J.~W.; Song,~B.~I.; Lee,~I.; Chung,~Y.~K. Dielectric Properties of PCB Embedded Bismuth-Zinc-Niobate Films Prepared by RF Magnetron Sputtering. 2008 10th Electronics Packaging Technology Conference. 2008; pp 1427--1430\relax
\mciteBstWouldAddEndPuncttrue
\mciteSetBstMidEndSepPunct{\mcitedefaultmidpunct}
{\mcitedefaultendpunct}{\mcitedefaultseppunct}\relax
\EndOfBibitem
\bibitem[Sharma \latin{et~al.}(2024)Sharma, Parasubotu, and Lodha]{PSedtm2024}
Sharma,~P.; Parasubotu,~Y.; Lodha,~S. High-k dielectric integration to improve breakdown characteristics of $\beta$-Ga$_2$O$_3$ Schottky diode. 2024 8th IEEE Electron Devices Technology I\& Manufacturing Conference (EDTM). 2024; pp 1--3\relax
\mciteBstWouldAddEndPuncttrue
\mciteSetBstMidEndSepPunct{\mcitedefaultmidpunct}
{\mcitedefaultendpunct}{\mcitedefaultseppunct}\relax
\EndOfBibitem
\bibitem[Qasrawi \latin{et~al.}(2014)Qasrawi, Muis, Rob, and Mergen]{qasrawi2014electrical}
Qasrawi,~A.; Muis,~K. O.~A.; Rob,~O. H. A.~A.; Mergen,~A. Electrical characterization of Bi$_{1.50-x}$Zn$_{0.92}$Y$_x$Nb$_{1.5}$O$_{6.92}$ varactors. \emph{Functional Materials Letters} \textbf{2014}, \emph{7}, 1450044\relax
\mciteBstWouldAddEndPuncttrue
\mciteSetBstMidEndSepPunct{\mcitedefaultmidpunct}
{\mcitedefaultendpunct}{\mcitedefaultseppunct}\relax
\EndOfBibitem
\bibitem[Sze \latin{et~al.}(2021)Sze, Li, and Ng]{sze2021physics}
Sze,~S.~M.; Li,~Y.; Ng,~K.~K. \emph{Physics of semiconductor devices}; John wiley \& sons, 2021\relax
\mciteBstWouldAddEndPuncttrue
\mciteSetBstMidEndSepPunct{\mcitedefaultmidpunct}
{\mcitedefaultendpunct}{\mcitedefaultseppunct}\relax
\EndOfBibitem
\bibitem[Pearton \latin{et~al.}(2018)Pearton, Yang, Cary, Ren, Kim, Tadjer, and Mastro]{PeartonAPR20182}
Pearton,~S.~J.; Yang,~J.; Cary,~I.,~Patrick~H.; Ren,~F.; Kim,~J.; Tadjer,~M.~J.; Mastro,~M.~A. {A review of Ga$_2$O$_3$ materials, processing, and devices}. \emph{Applied Physics Reviews} \textbf{2018}, \emph{5}, 011301\relax
\mciteBstWouldAddEndPuncttrue
\mciteSetBstMidEndSepPunct{\mcitedefaultmidpunct}
{\mcitedefaultendpunct}{\mcitedefaultseppunct}\relax
\EndOfBibitem
\bibitem[Konishi \latin{et~al.}(2017)Konishi, Goto, Murakami, Kumagai, Kuramata, Yamakoshi, and Higashiwaki]{Konishi2017}
Konishi,~K.; Goto,~K.; Murakami,~H.; Kumagai,~Y.; Kuramata,~A.; Yamakoshi,~S.; Higashiwaki,~M. 1-kV vertical Ga$_2$O$_3$ field-plated Schottky barrier diodes. \emph{Applied Physics Letters} \textbf{2017}, \emph{110}\relax
\mciteBstWouldAddEndPuncttrue
\mciteSetBstMidEndSepPunct{\mcitedefaultmidpunct}
{\mcitedefaultendpunct}{\mcitedefaultseppunct}\relax
\EndOfBibitem
\bibitem[Labed \latin{et~al.}(2023)Labed, Min, Jo, Sengouga, Venkata~Prasad, and Rim]{FLP2023reduction}
Labed,~M.; Min,~J.~Y.; Jo,~E.~S.; Sengouga,~N.; Venkata~Prasad,~C.; Rim,~Y.~S. Reduction of Fermi-Level Pinning and Controlling of Ni/$\beta$-Ga2O3 Schottky Barrier Height Using an Ultrathin HfO2 Interlayer. \emph{ACS Applied Electronic Materials} \textbf{2023}, \emph{5}, 3198--3205\relax
\mciteBstWouldAddEndPuncttrue
\mciteSetBstMidEndSepPunct{\mcitedefaultmidpunct}
{\mcitedefaultendpunct}{\mcitedefaultseppunct}\relax
\EndOfBibitem
\bibitem[Dhara \latin{et~al.}(2022)Dhara, Kalarickal, Dheenan, Joishi, and Rajan]{dhara2022beta}
Dhara,~S.; Kalarickal,~N.~K.; Dheenan,~A.; Joishi,~C.; Rajan,~S. $\beta$-Ga$_2$O$_3$ Schottky barrier diodes with 4.1 MV/cm field strength by deep plasma etching field-termination. \emph{Applied Physics Letters} \textbf{2022}, \emph{121}\relax
\mciteBstWouldAddEndPuncttrue
\mciteSetBstMidEndSepPunct{\mcitedefaultmidpunct}
{\mcitedefaultendpunct}{\mcitedefaultseppunct}\relax
\EndOfBibitem
\bibitem[He \latin{et~al.}(2021)He, Hao, Zhou, Li, Zhou, Chen, Xiong, Jian, Xu, Zhao, \latin{et~al.} others]{he2021over}
He,~Q.; Hao,~W.; Zhou,~X.; Li,~Y.; Zhou,~K.; Chen,~C.; Xiong,~W.; Jian,~G.; Xu,~G.; Zhao,~X.; others Over 1 GW/cm$^2$ vertical Ga$_2$O$_3$ Schottky barrier diodes without edge termination. \emph{IEEE Electron Device Letters} \textbf{2021}, \emph{43}, 264--267\relax
\mciteBstWouldAddEndPuncttrue
\mciteSetBstMidEndSepPunct{\mcitedefaultmidpunct}
{\mcitedefaultendpunct}{\mcitedefaultseppunct}\relax
\EndOfBibitem
\bibitem[Sasaki \latin{et~al.}(2017)Sasaki, Wakimoto, Thieu, Koishikawa, Kuramata, Higashiwaki, and Yamakoshi]{sasaki2017first}
Sasaki,~K.; Wakimoto,~D.; Thieu,~Q.~T.; Koishikawa,~Y.; Kuramata,~A.; Higashiwaki,~M.; Yamakoshi,~S. First demonstration of Ga$_2$O$_3$ trench MOS-type Schottky barrier diodes. \emph{IEEE Electron Device Letters} \textbf{2017}, \emph{38}, 783--785\relax
\mciteBstWouldAddEndPuncttrue
\mciteSetBstMidEndSepPunct{\mcitedefaultmidpunct}
{\mcitedefaultendpunct}{\mcitedefaultseppunct}\relax
\EndOfBibitem
\end{mcitethebibliography}


\end{document}